\documentclass[final,5p,times,twocolumn,authoryear]{elsarticle}
\biboptions{numbers,sort&compress}

\usepackage{amsmath}
\usepackage{amssymb}
\usepackage{lipsum}
\usepackage[T1]{fontenc}
\usepackage{url}
\usepackage{natbib}
\setcitestyle{numbers,square,comma}

\usepackage{caption}
\usepackage{floatrow}

\usepackage[para,online,flushleft]{threeparttablex}

\usepackage{hyperref}
\hypersetup{breaklinks=true,colorlinks=true,linkcolor=blue,citecolor=blue,filecolor=magenta,urlcolor=cyan}
\newcommand{\red}[1]{ {#1} }
\usepackage[all]{hypcap}

\usepackage{tabularx}
\usepackage{xcolor}
\definecolor{pastelgray}{rgb}{0.81, 0.81, 0.77}
\definecolor{beaublue}{rgb}{0.9, 0.9, 0.93}

\renewcommand{\vec}[1]{\mbox{\boldmath $#1$}}

\journal{Physics Letters B}

\begin{document}

\begin{frontmatter}



\title{Exploring the possible two-proton radioactivity of $^{38,39}$Ti}


\author[ad1,ad2]{B. Huang}
\author[ad1,ad2]{F. P. Bai}
\author[ad3]{J. G. Li}
\author[ad1,ad2]{D. Q. Fang\corref{mailaddress1}}
\author[ad1,ad2]{S. M. Wang\corref{mailaddress2}}

\address[ad1]{Key Laboratory of Nuclear Physics and Ion-beam Application (MOE), Institute of Modern Physics, Fudan
University, Shanghai 200433, China}
\address[ad2]{Shanghai Research Center for Theoretical Nuclear Physics,
NSFC and Fudan University, Shanghai 200438, China}
\address[ad3]{Institute of Modern Physics, Chinese Academy of Sciences, Lanzhou 730000, China}

\cortext[mailaddress1]{dqfang@fudan.edu.cn}
\cortext[mailaddress2]{wangsimin@fudan.edu.cn}

\begin{abstract}
Two-proton (2$p$) radioactivity represents a rare decay mode that has been experimentally observed only in a selected few nuclei. The exploration of 2$p$ emission is crucial for elucidating the structure, mass, and nucleon-nucleon interactions within exotic proton-rich nuclei. $^{39}$Ti has long been postulated as a potential candidate for 2$p$ emission; however, experimental investigations have yet to confirm its 2$p$ decay. To provide more accurate information for further studies, we utilize the Gamow shell model (GSM) and the Gamow coupled channel (GCC) method to analyze the prospective 2$p$ radioactivity of isotopes $^{38,39}$Ti. Our calculations suggest that $^{39}$Ti is indeed a viable candidate for 2$p$ emission. Notably, the estimated partial 2$p$ decay width for $^{39}$Ti, predicted from the three-body GCC method, suggests that its 2$p$ decay could rival its $\beta$ decay in likelihood, although this is highly dependent on the specific 2$p$ decay energy. Additionally, our analysis indicates a propensity for pairing between the valence protons in $^{39}$Ti. A similar investigative approach reveals that $^{38}$Ti exhibits a higher 2$p$ decay energy and a broader decay width than $^{39}$Ti, positioning it as a more promising candidate for 2$p$ decay. 
\end{abstract}




\end{frontmatter}




\section{Introduction}
\label{introduction}

Two-proton (2$p$) decay, an exotic nuclear decay mode, was first proposed in the 1960s by Goldansky~\citep{GOLDANSKY1960a,GOLDANSKY1961a}. This decay process is characteristic of some unbound even-$Z$ nuclei located beyond the proton drip line, where spontaneous emission of two protons occurs. Early theoretical efforts provided a foundational description of 2$p$ emission, predicting potential candidates for experimental verification~\cite{GOLDANSKY1961a,GALITSKY1964a,JANECKE1965a}, including $^{39}$Ti, $^{42}$Cr, $^{45}$Fe, and $^{48,49}$Ni~\cite{Brown1991a,Brown1991b,Cole1996a,Ormand1996a}. 
Initial experimental investigations aimed to detect 2$p$ decay in lighter nuclei such as $^{6}$Be, $^{12}$O, and $^{16}$Ne, which are relatively easier to produce~\cite{Geesaman1977a,BOCHKAREV1989a,KeKelis1978a,Kryger1995a}. However, these nuclei exhibit very short half-lives and broad intermediate states, complicating the differentiation between simultaneous and sequential 2$p$ emissions. A similar challenge was encountered in studies of the 2$p$ decay of $^{19}$Mg~\cite{Mukha2007a,Mukha2008a}. 
The discovery of long-lived ground state (g.s.) 2$p$ emitters occurred significantly later. Since the first direct observation of 2$p$ radioactivity in the g.s. of $^{45}$Fe in 2002, this decay mode has been documented in only four medium-heavy nuclides: $^{45}$Fe~\cite{Giovinazzo2002a,Pfutzner2002a}, $^{54}$Zn~\cite{Blank2005a,Ascher2011a}, $^{48}$Ni~\cite{Dossat2005a,Pomorski2011a}, and $^{67}$Kr~\cite{Goigoux2016a,Grigorenko2017a,Wang2018a}. 
Given the critical insights that studies of 2$p$ decay provide on the structure of proton-rich nuclei and nucleon-nucleon correlations, the quest to identify more 2$p$ decay candidates through both experimental and theoretical research continues to be a focal area of intense academic interest~\cite{Pfutzner2012a,Blank2008a,Grigorenko2009a,Zhou2022a,Pfutzner2023a}.

$^{39}$Ti, as a potential candidate for long-lived g.s. 2$p$ emission, has garnered significant attention within the nuclear physics community~\cite{GOLDANSKII1988a}. Experimental investigations for $^{39}$Ti have been conducted on multiple occasions. Initially identified in 1990, $^{39}$Ti did not exhibit direct signs of 2$p$ decay at that time~\cite{DETRAZ1990a}, with its half-life measured at $26_{-7}^{+8}$ ms. Subsequent research in 1992 revealed the beta-delayed two-proton ($\beta$2$p$) decay of $^{39}$Ti~\cite{Moltz1992a}. Further experimental efforts in 2001 and 2007 refined the half-life measurements to $31_{-4}^{+6}$ ms and $28.5_{-9}^{+9}$~ms, respectively~\cite{Giovinazzo2001a,DOSSAT2007a}. Despite these detailed investigations, neither experiment succeeded in observing the g.s. 2$p$ emission of $^{39}$Ti.

The theoretical description of 2$p$ decay in proton-rich nuclei presents a formidable challenge to contemporary theoretical frameworks, chiefly due to the necessity to comprehensively model the inner nuclear structure, few-body asymptotic behavior, and continuum effects. Due to the intricate nature of these phenomena, existing theoretical models tend to focus on one or two of these aspects, often leaving a gap in a comprehensive theoretical representation \cite{Grigorenko2009a,Pfutzner2023a,Zhou2022a}. Notably, several approaches have been applied to the study of 2$p$ emission from $^{39}$Ti. The mass relations of mirror nuclei and isospin multiplets have been instrumental in estimating the two-proton separation energy for $^{39}$Ti, suggesting its potential for exhibiting 2$p$ radioactivity~\cite{Brown1991a,Ormand1996a,Cole1997a,Grigorenko2001a,Tian2013a,MC2020a}. 
Moreover, based on the predicted two-proton separation energy, both the extended $R$-matrix framework \cite{ZHANG2023a} and the semi-classical Wentzel-Kramers-Brillouin (WKB) approximation method \cite{Cui2020a,Royer2022a,Santhosh2021a,Santhosh2022a} have been utilized to estimate the 2$p$ radioactivity half-life of $^{39}$Ti. Bayesian statistical tools have further highlighted $^{39}$Ti as a prime candidate for 2$p$ decay \cite{Neufcourt2020a}. 
Meanwhile, the neighboring isotope, $^{38}$Ti, which remains experimentally undiscovered, has been the subject of extensive theoretical studies since the 1990s. Numerous models have assessed and underscored its potential for exhibiting 2$p$ decay \cite{ZHANG2023a,Mehana2023a,Neufcourt2020a,Royer2022a,XU2006a}. This ongoing theoretical interest highlights the importance of $^{38,39}$Ti in expanding our understanding of nuclear decay processes at the proton dripline. It is important to recognize that these studies, while pioneering, have not provided a detailed and microscopic analysis of the decay properties and continuum effects concerning $^{38,39}$Ti. Therefore, further research is essential to enhance our understanding of these isotopes and other systems at the nuclear dripline.

For unbound 2$p$ emitters situated far from the stability line, the effects of the continuum are pronounced~\cite{Okolowicz2012a, Michel_2021b}. To effectively integrate the internal nuclear structure with continuum coupling effects, various theoretical frameworks have been developed. These include the Shell Model Embedded in the Continuum (SMEC)~\cite{Bennaceur2000a, Okolowicz2003a}, the Gamow Shell Model (GSM)~\cite{Betan2002a, Michel2002a, Forssen2013a, Jaganathen2017a}, and the Gamow Coupled-Channel Method (GCC)~\cite{Wang2017a, Wang2018a}. SMEC and GSM primarily focus on the configuration mixing influenced by continuum interactions~\cite{Rotureau2005a, Michel2021a}, whereas GCC is tailored to analyze few-body decay properties~\cite{Wang2019a, Yang2023a}. These methodologies have been effectively employed to investigate the structures~\cite{Michel2021a}, half-lives~\cite{Rotureau2005a, Rotureau2006a}, and decay mechanisms~\cite{Wang2021a, Wang2022a} of 2$p$ emitters. 
In our current study, we employ both GSM and GCC to provide a comprehensive investigation of the structural and decay properties of the isotopes $^{38,39}$Ti, aiming to enhance our understanding of their behavior within the context of nuclear decay phenomena far from stability.

This article is structured as follows: Section II delineates the theoretical frameworks of GSM and GCC, detailing the specific forms of effective interactions employed in this study. Section III provides a comprehensive presentation of the calculated spectra and half-lives for isotopes $^{38,39}$Ti, including the 2$p$ density distribution for $^{39}$Ti. Section IV summarizes the principal findings and conclusions derived from this research.

\section{Method}
The GSM extends the traditional shell model into an open quantum system framework~\cite{Betan2002a,Michel2002a,SUN2017a}, utilizing a one-body Berggren basis generated by a finite-depth potential~\cite{BERGGREN1968a}. This basis is defined in the complex plane, adhering to the completeness relation: 
\begin{equation}\label{Berggen}
    \sum\limits_n {\left| {{u_n}} \right\rangle } \left\langle {{{\tilde u}_n}} \right| + \int_{{L_ + }} {\left| {{u_k}} \right\rangle } \left\langle {{{\tilde u}_k}} \right|dk = 1,
\end{equation}
where ${u_n}$ symbolizes bound or resonance states, and ${u_k}$ denotes the continuum states along the contour ${L+}$ in the complex plane~\cite{BERGGREN1968a}. This formulation allows GSM to provide a comprehensive description of both structural and decay properties within a many-body framework, thereby positioning it as an essential tool for predicting the properties of nuclei near or beyond the drip line. In practical calculations, the Gauss-Legendre quadrature method is utilized to discretize the ${L_+}$ contour in Eq.\,(\ref{Berggen}), the ${L_ + }$ contour follows the path $k = 0 \to 0.2 - 0.15i \to 0.4 \to 4$ (all in ${\rm{f}}{{\rm{m}}^{ - 1}}$), where each segment is discretized by 10 points along the ${L_ + }$ contour.~ {The selected integration contour and its discretization have been verified to have good stability. The changes of the spectra are within 30\,keV for different contours and more discretization points.} The adoption of the Berggren basis in these calculations is pivotal for accurately capturing nuclear correlations through configuration mixing.

In this study, we model the nucleus using a core-plus-valence-protons framework, based on the cluster-orbital shell model coordinate. We specifically select the stable nucleus $^{36}$Ca as the core for this investigation. The one-body potential is described by a Woods-Saxon (WS) potential. Meanwhile, the two-body interaction is represented through a Effective Field Theory (EFT) approach, augmented by a nucleon-number-dependent factor, which serves to emulate the effects typically attributed to the three-body forces, as evidenced by the enhancement in the replication of experimental energy levels~\cite{MACHLEIDT2011a,CONTESSI2017a,Hammer2013a,Hammer2020a}. The exponential parameter for the $A$-dependent two-body factor is set at 0.3, consistent with parameters from other investigations~\cite{Michel2019a,Brown2006a,Huth2018a}. This adaptation of the GSM with EFT interactions has proven instrumental in exploring the properties of proton-rich nuclei around $A \approx 20$, as well as neutron-rich isotopes of oxygen~\cite{Michel2019a,Li2021b}. Typically, EFT interactions require renormalization through proper regulators.  {In the current framework, the EFT interaction is renormalized as described in Ref.~\cite{MACHLEIDT2011a}, using a momentum-dependent regulator function $f(p,p^\prime)$ defined as follows:
\begin{eqnarray}
V_{\rm EFT}(\vec{p}^\prime,\vec{p}) &\rightarrow& V_{\rm EFT}(\vec{p}^\prime,\vec{p})f(p^\prime,p) \label{EFT_renormalization} \\
f(p^\prime,p) &=& \exp \left[ -\left( \frac{p^\prime}{\Lambda} \right)^{2n} - \left( \frac{p}{\Lambda} \right)^{2n} \right] \label{EFT_regulator}
\end{eqnarray}
where $\Lambda$ represents a cut-off energy of 300 MeV, and $n=2,3,4$ depending on the LO or NLO constant and a two-nucleon partial wave is considered.} Recent studies affirm that EFT interactions, when regularized within the confines of a Harmonic Oscillator (HO) basis, yield convergent and reliable results for heavier nuclei, particularly within a rigorously defined model space~\cite{Huth2018a,Binder2016a,Bansal2018a}.

In the regular shell-model framework, the Fermi surfaces for $Z = 22$ and $N = 17$ in $^{39}$Ti are primarily $0f_{7/2}$ and $0d_{3/2}$ orbitals, respectively. The configurations, wherein nucleons occupy the $sd$ and $pf$ continua, are considered active components in our analysis. Consequently, we have selected the $sdpf$ model space for our GSM calculations.  {In the current framework, mixed bases, including the HO and Berggren ensembles, are employed. For the valence protons, the $p_{1/2}$ and $p_{3/2}$ partial waves are represented using the Berggren basis, while the $f_{5/2}$ and $f_{7/2}$ partial waves are calculated using the HO basis.} Due to the relatively large centrifugal barrier, the $f$-wave experiences a small interplay with the continuum, which justifies its representation via the HO basis. This choice effectively reduces the dimensions of the model space and the computational demand. For valence neutrons, the orbitals $0d_{3/2}$, $0f_{7/2}$, and $1p_{3/2}$ are bound, hence similarly expressed using the HO basis. Parameters for the one-body WS potential, such as the diffuseness $d$ and the spin-orbit coupling strength $V_{ls}$, are aligned with the values reported in Ref.~\cite{Jaganathen2017a}. The radius parameter $R_0$ is set to 3.72 fm. Additionally, the depth $V_0$ for each orbital has been adjusted to reproduce the experimental energies for the ground and low-lying excited states of isotopes $^{37}$Ca and $^{39}$Sc. The depths for the proton $s$, $p$, $d$, and $f$ partial waves are set at 63.42 MeV, 66.38\,MeV, 63.42 MeV, and 61.55~MeV, respectively, while the corresponding depths for neutrons are 62.92\,MeV, 62.92\,MeV, 63.82\,MeV, and 67.38~MeV.

\begin{table}[!htb]
\caption{Optimized parameters of the EFT-inspired interaction at leading order (LO) and next-to-leading order (NLO) in natural units. Detailed definitions and symbols can be found in  {Supplemental Material (SM) \cite{SM} and} Refs.~\cite{CONTESSI2017a,Hammer2013a,Hammer2020a}.}
\begin{tabular}{cccc}
\hline\hline\vspace{-0.2cm} \\
LO constant & LO value & NLO constant & NLO value   \\ 
\hline \vspace{-0.2cm} \\ 
$C_{^{1}S_0}$ & $-0.77$ & $C_1$ & $-1.23$ \\\vspace{-0.2cm} \\
$C_{^{3}S_1}$ & $-22.13$ & $C_2$ & 2.74 \\\vspace{-0.2cm} \\
$C_{^{1}T_0}$ & 0 & $C_3$ & $-1.23$ \\\vspace{-0.2cm} \\
$C_{^{3}T_1}$ & $-3.60$ & $C_4$ & $-1.45$ \\\vspace{-0.2cm} \\
&& $C_5$ & $-2.31$ \\\vspace{-0.2cm} \\
&& $C_6$ & $-0.66$ \\\vspace{-0.2cm} \\
&& $C_7$ & $-2.45$ \\\vspace{-0.2cm} \\
\hline\hline
\end{tabular}\label{Table.InterParamEFT}
\end{table}

Table~\ref{Table.InterParamEFT} presents the optimized parameters of the  {EFT-inspired} interaction utilized in this study. The EFT-inspired interaction incorporates both leading-order components, denoted as $C_{^{1}S_0}$, $C_{^{3}S_1}$, $C_{^{1}T_0}$, and $C_{^{3}T_1}$, and next-to-leading order terms, represented by ${C_{1...7}}$. $C_{^{1}S_0} = (C_S-3C_{_T})$ and $C_{^{3}S_1}= (C_S+C_{_T})$ are interaction strengths for the $^1S_0$ and $^3S_1$ channels, respectively. The constants ${C_S}$ and ${C_T}$ are contingent upon the isospin $T = 0,1$ of the interacting nucleons~\cite{CONTESSI2017a,Hammer2013a,Hammer2020a}. In our specific theoretical setup, the multiplicity of these constants is consolidated, allowing for the simplification to a singular optimized constant, ${C_S}$.  {The optimization of the interactions is carried out following the approach outlined in Ref.~\cite{Dobaczewski2014a}. In the optimization model, optimized constant ${C_S}$ are tuned to match 13 observables $\mathcal{O}_i$ (where $i = 1, \dots, 13$), which are binding energies of 13 states of Ca, Sc, and Ti isotope chains near $^{39}$Ti. The optimization process is based on minimizing the penalty function:
\begin{equation}
\chi^2(\vec{C_S}) = \sum_{i=1}^{13}\left(\frac{\mathcal{O}_i(\vec{C_S}) - \mathcal{O}_i^{\mbox{\scriptsize{exp}}}}{\delta \mathcal{O}_i} \right) ^2,\label{chiFunction}
\end{equation}
where $\mathcal{O}_i(\vec{C_S})$ are the model-calculated observables. The $\mathcal{O}_i^{\mbox{\scriptsize{exp}}}$ are the corresponding experimental data (fit-observables) used to constrain the model. The errors $\delta \mathcal{O}_i$ account for various contributions from experimental uncertainties, numerical inaccuracies, and theoretical errors due to model limitations. More details of the optimization method are supplemented in SM \cite{SM}.}

To explore the few-body decay properties of isotopes $^{38,39}$Ti, GCC method, as described in Refs.~\cite{Wang2017a,Wang2018a}, is employed for validation and as a supplementary analysis to GSM. The GCC method, like GSM, adopts a complex-plane approach utilizing the Berggren basis but distinguishes itself by implementing a three-body model using Jacobi coordinates.  {Commonly used Jacobi coordinates include Jacobi-$T$ and -$Y$ coordinates are illustrated in SM \cite{SM}.} This model configuration effectively manages both the center-of-mass motion and the asymptotic behaviors of the system. Consequently, the GCC method is adept at capturing the correlations between valence nucleons as well as accurately computing the three-body decay widths.

In the GCC framework, the system is described as a core plus two valence nucleons or clusters. The Hamiltonian of this configuration is expressed as:
\begin{equation}
    \hat{H} = \sum^3_{i=1}\frac{ \hat{\vec{p}}^2_i}{2 m_i} +\sum^3_{i>j=1} V_{ij}(\vec{r}_{ij})-\hat{ T}_{\rm c.m.},
\end{equation}
where $V_{ij}$ represents the interaction between clusters $i$ and $j$, and $\hat{T}_{\rm c.m.}$ denotes the kinetic energy associated with the center-of-mass. The wave functions of the valence protons are expressed in Jacobi coordinates, facilitating a precise representation of three-body asymptotics and elimination of spurious center-of-mass movements. 

For the core-proton (core-$p$) effective interaction, a WS potential  { with ``universal'' parameters~\cite{Cwiok1987a} is utilized. These parameters were initially proposed to study the high-spin states of heavy-mass nuclei \cite{Dudek1982}, and have further been tested in Ref.\,\cite{Nazarewicz1985} for light nuclei showing satisfactory performance in describing the single (quasi) particle level sequences. Meanwhile, the half-life or decay width of a system is highly sensitive to the decay energy. Therefore, to provide more accurate estimates of the decay properties of the nuclei of interest,} the depth $V_{0}$ is specifically tuned to align the g.s. energy of $^{38}$Ti or $^{39}$Ti with the results obtained from GSM. The nuclear two-body interaction between valence nucleons is using the finite-range Minnesota force, with original parameters sourced from Ref.~\cite{Thompson1977a}. {This set of parameters is derived from a fit to experimental data including the scattering phase shift of $n$ + $^{40}$Ca, and has been shown to provide an accurate description of the properties of nuclides in the region surrounding $^{40}$Ca}~\cite{Thompson1977a,ENDT1978a}.

In this study, the Berggren basis was implemented for channels with $K \leqslant 5$ within the GCC framework, while the HO basis was adopted for channels associated with higher angular momentum. The complex-momentum contour for the Berggren basis in the GCC method is defined as $k = 0 \to 0.3 - 0.1i \to 0.4 - 0.05i \to 0.5 \to 0.8 \to 1.2 \to 2 \to 4 \to 6$ fm$^{-1}$, with each segment discretized into 20 scattering states. The HO basis is set with an oscillator length $b = 1.75$ fm and a maximum principal quantum number $n_{\rm max} = 20$.

\section{Results and discussion}

\begin{figure*}[htb]
\floatbox[{\capbeside\thisfloatsetup{capbesideposition={right,top},capbesidewidth=4cm}}]{figure}[\FBwidth]
{\caption{Spectra of Ca, Sc, and Ti isotope chains near $^{39}$Ti, calculated using the optimized GSM interaction. Experimental values are represented by stars, as taken from~\cite{NNDC}.}\label{figreferencelevel}}
{\includegraphics[width=0.75\textwidth]{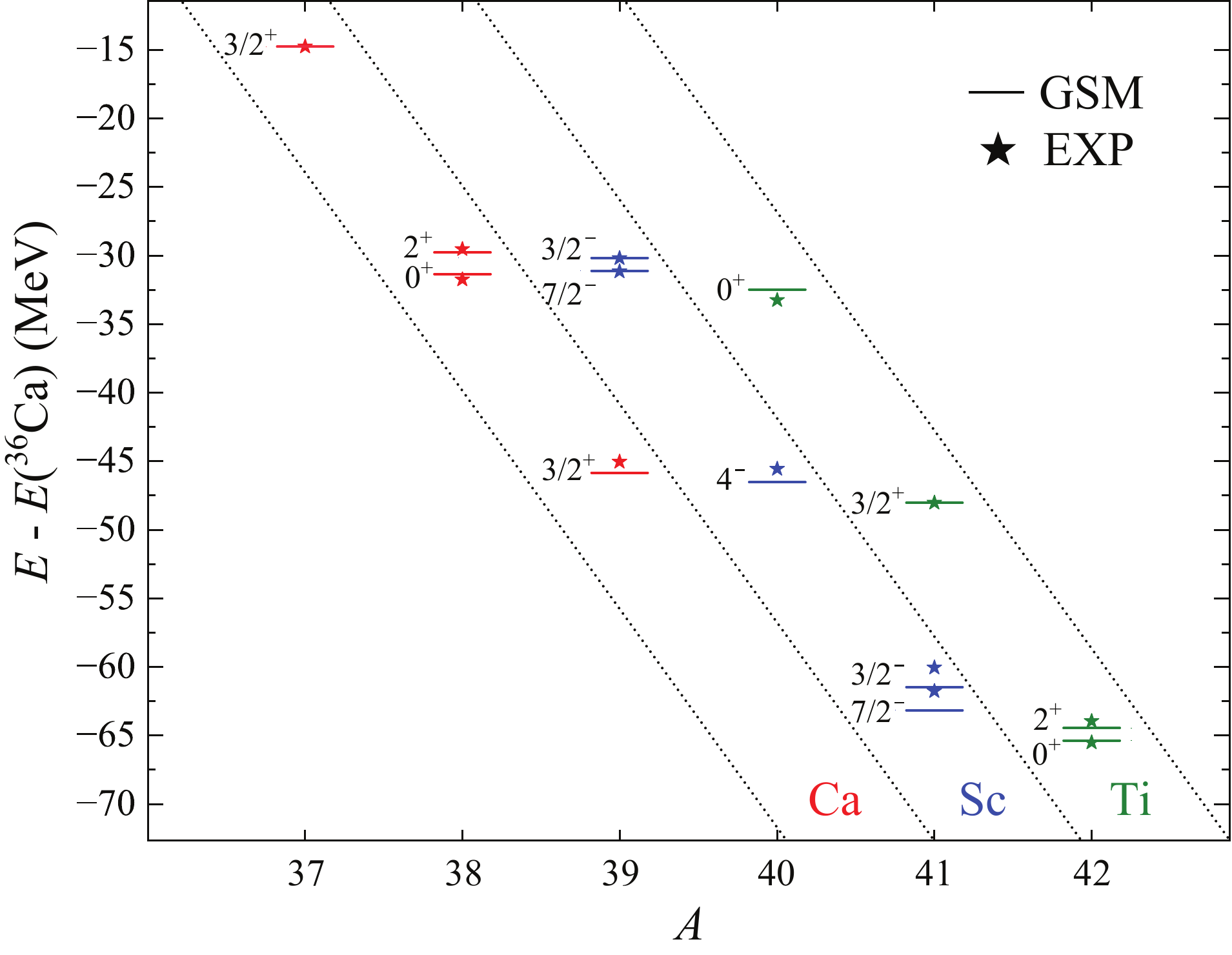}}
\end{figure*}

The results of the optimization are presented in Fig.~\ref{figreferencelevel}, where the GSM calculations were performed using the optimized EFT-inspired interactions. It displays a commendable concordance with the experimental spectra of nuclides around the proton drip-line. The overall root mean square deviation (RMSD) for these reference states is quantified as follows:
\begin{equation}\label{RMSD}
    {\rm RMSD}=\sqrt {\sum\limits_{i = 1}^{13} {{{\left( {{E^{\rm GSM}_{i}} - {E^{\rm EXP}_{i}}} \right)}^2}} /13},
\end{equation}
where $E^{\rm GSM}_{i}$ represents the GSM-predicted energy for the $i$-th reference state among the selected 13 states, and $E^{\rm EXP}_{i}$ denotes the corresponding experimental energy. The RMSD is approximately 712 keV. The primary sources of deviation can be attributed to the  {open-shell nature} of the core $^{36}$Ca, as noted by Ref.~\cite{Dronchi2023a}, and to minor discrepancies observed in nuclei distant from the stability line.

\begin{figure}
	\centering 
	\includegraphics[width=1\columnwidth]{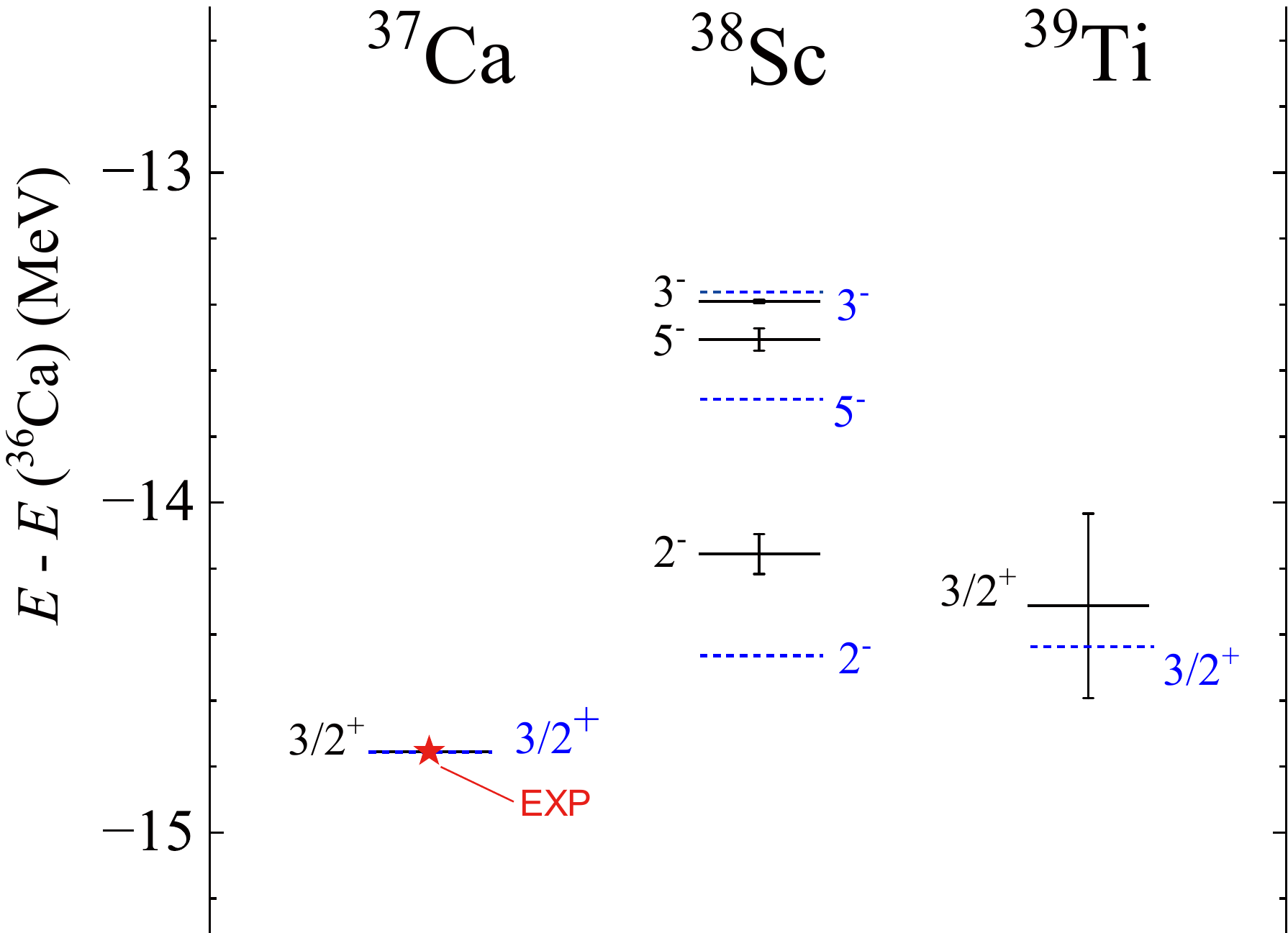}	
	\caption{Predicted energy spectra for $^{37}$Ca, $^{38}$Sc, and $^{39}$Ti, alongside the estimated uncertainties for each level. Additionally, the experimental values denoted by stars are taken from Ref.~\cite{NNDC}.  {As a comparision, the results with the channel-independent WS potential are calculated as shown by the blue dashed lines.} Since the g.s. energy of $^{37}$Ca is incorporated within the optimization parameters, the corresponding uncertainty for this nucleus is omitted.} 
	\label{fig_39Ti}%
\end{figure}

The calculated spectra for the isotones $^{37}$Ca, $^{38}$Sc, and $^{39}$Ti are shown in Fig.~\ref{fig_39Ti}. To estimate the uncertainties associated with the predicted energies, we readjusted the strength of the nucleon-nucleon interaction by $2\%$  {using a global scaling factor}, and the resulting uncertainties are represented by error bars in the same figure. These uncertainties do not significantly influence the internal structure of the nuclei, but can affect the relative energy spectra among them.
The GSM calculations reveal that the $3/2^+$ g.s. energies for $^{39}$Ti and $^{37}$Ca are $-14.31$\,MeV and $-14.76$\,MeV, respectively. Meanwhile, the neighboring nucleus $^{38}$Sc exhibits a $2^-$ g.s. energy of $-14.16$\,MeV, alongside $5^-$ and $3^-$ excited states. Consequently, the two-proton separation energy ($S_{2p}$) for $^{39}$Ti is $-0.45$\,MeV, and the single-proton separation energy ($S_p$) is $0.15$\,MeV. This level-arrangement permits the 2$p$ decay of $^{39}$Ti, while energetically prohibiting single-proton decay, aligning with findings from previous studies \cite{Brown1991a,Ormand1996a,Tian2013a,Neufcourt2020a,MC2020a}.
Furthermore, the uncertainties in the g.s. energies for $^{39}$Ti and $^{38}$Sc are 0.3\,MeV and 0.06\,MeV, respectively.  {Considering these uncertainties, $S_{2p}$ for $^{39}$Ti is $-0.45$\,MeV, with an uncertainty of 0.3 MeV, which makes $S_{2p}$ remains negative. This suggests the potential for 2$p$ decay of $^{39}$Ti. Additionally, the $S_{p}$ of $^{39}$Ti is $0.15$\,MeV, with an uncertainty of 0.31 MeV}. Hence, the 2$p$ decay for $^{39}$Ti could manifest as either direct 2$p$ decay or through a sequential emission process, affecting the half-life or branching ratio of this decay channel.
To ascertain the primary decay channels for $^{39}$Ti, we analyzed the configuration information for its $3/2^+$ g.s. The dominant configuration is represented as ${}^{36}{\rm{Ca}} \otimes \pi \left( {f_{7/2}} \right)^2\nu \left( {d_{3/2}} \right)^1$, with an occupation probability of approximately $63\%$. A secondary, yet significant, configuration is ${}^{36}{\rm{Ca}} \otimes \pi \left( {p_{3/2}} \right)^2\nu \left( {d_{3/2}} \right)^1$, with about $22\%$ probability. These configurations indicate that the valence protons predominantly occupy $p$-wave and $f$-wave orbitals. Notably, the $3/2^-$ and $7/2^-$ levels are lying closely in isotopes like $^{39}$Sc or $^{41}$Sc, which suggests a competitive interaction between $p$-wave and $f$-wave orbitals in this region. The $f$-wave is predominant in the g.s. of $^{39}$Ti, while the $p$-wave significantly influences the decay process due to its lower centrifugal barrier.

 {In addition, to estimate the uncertainty caused by the single-particle potentials, the spectra of $^{38}$Sc and $^{39}$Ti (see the dashed levels in Fig.~\ref{fig_39Ti}) are also calculated based on a channel-independent WS potential, where the potential depths of each channel for proton and neutron were fixed at 62.12 MeV and 63.82 MeV, respectively. The potential depths for neutron and proton are adjusted by fitting the experimental energies of the ground and low-lying excited states of $^{37}$Ca and $^{39}$Sc. In this case, the resulting g.s. energies of $^{39}$Ti, $^{38}$Sc, and $^{37}$Ca are $-14.44$\,MeV, $-14.46$\,MeV, and $-14.76$\,MeV, respectively, and the corresponding $S_{2p}$ and $S_{p}$ of $^{39}$Ti are $-0.32$ MeV and $-0.02$\,MeV, respectively. This result is consistent with the previous prediction that $^{39}$Ti may have 2$p$ radioactivity.}

\begin{figure}
	\centering 
	\includegraphics[width=1\columnwidth, angle=0]{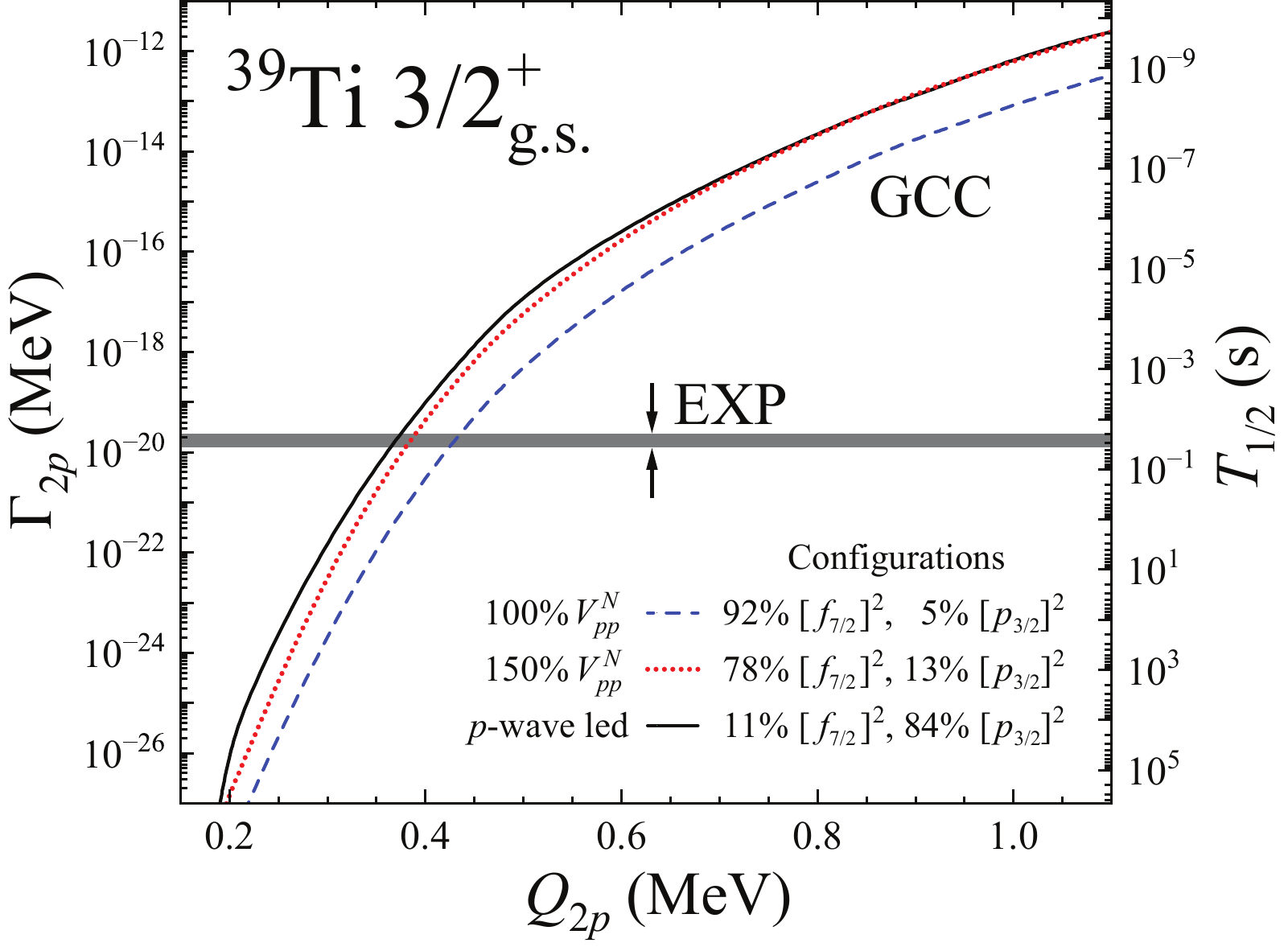}	
	\caption{The calculated 2$p$ partial width (half-life) for the g.s. of $^{39}$Ti as a function of the 2$p$ decay energy $Q_{2p}$. The Minnesota force is utilized for $V_{pp}^N$, considering interaction strengths of $100\%$ (dashed and solid lines) and $150\%$ (illustrated with a dotted line). To refine the estimate of the decay width limit, the single-particle energies were adjusted to enhance the contribution from the $p$-wave component, which is depicted as a solid line. The experimental data, represented by a shaded band, are sourced from Refs.~\cite{DETRAZ1990a,Giovinazzo2001a,DOSSAT2007a}.
} 
	\label{fig_39Tidecaywidth}%
\end{figure}

\begin{figure*}
	\centering 
	\includegraphics[width=1.0\textwidth, angle=0]{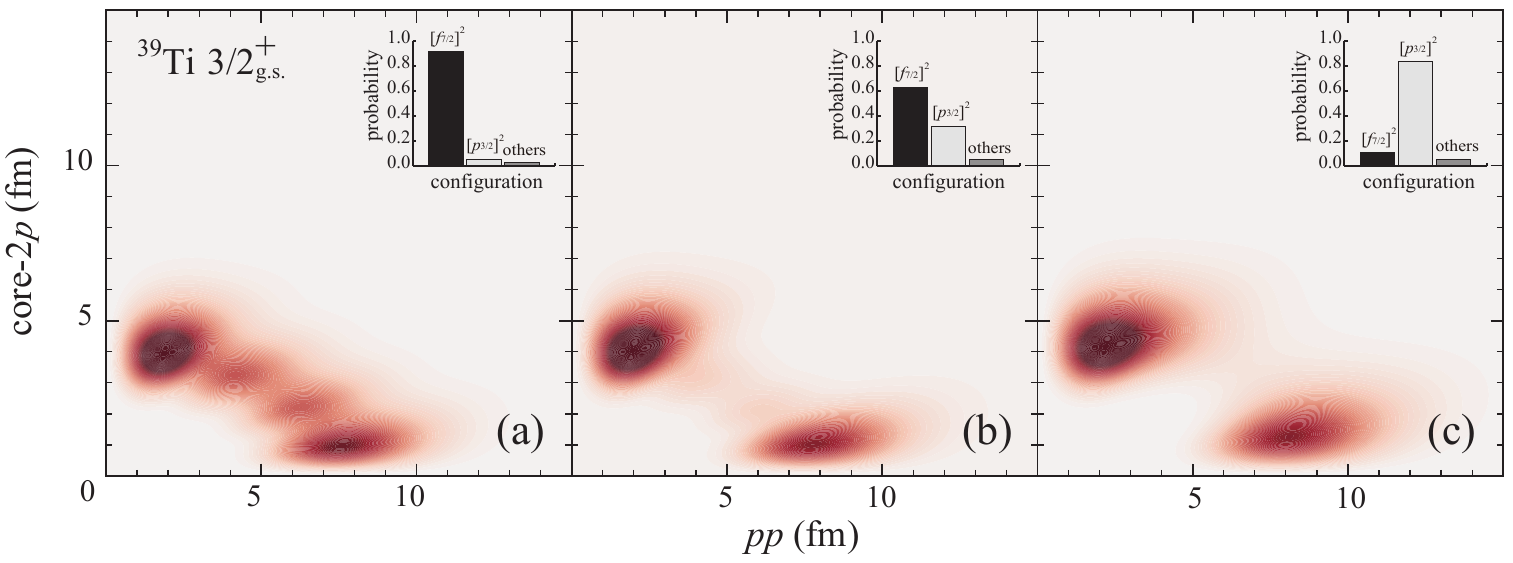}	
	\caption{Calculated 2$p$ density distributions with different configurations for the g.s. of $^{39}$Ti in Jacobi coordinates $pp$ and core-$2p$.} 
	\label{fig_39Ti_2$p$density}%
\end{figure*}

From the analysis of the decay energy of $^{39}$Ti, it is evident that three-body decay mechanisms may be involved. In contrast to the GSM, the three-body GCC method incorporates Jacobi cooridinates and captures the three-body decay lifetime with greater accuracy~\cite{Wang2018a}. Additionally, the GCC calculations include configurations of positive-parity single-particle states for valence nucleons, thereby facilitating a more detailed investigation of the cross-shell effects on valence protons. Thus, the GCC method is also applied in this study to examine the 2$p$ decay lifetime and configurations of $^{39}$Ti.
The partial 2$p$ decay width of the g.s. of $^{39}$Ti demonstrates high sensitivity to decay energy, as illustrated in Fig.~\ref{fig_39Tidecaywidth}. Here, a mere 100 keV difference in decay energy can precipitate a change in the half-life or decay width by approximately 1 to 5 orders of magnitude. To gauge the uncertainty in the calculated half-life, we considered the effects of proton-proton interactions and configurations. The dashed blue and dotted red lines in Fig.~\ref{fig_39Tidecaywidth} correspond to scenarios where the Coulomb force remains constant, and the strength of the proton-proton nuclear force $V_{pp}^N$ is at $100\%$ and $150\%$, respectively. Notably, a $50\%$ increase in $V_{pp}^N$ leads to an order of magnitude increase in the predicted ${\Gamma _{2p}}$, indicative of the facilitative role of proton pairing in $^{39}$Ti for the tunneling process, suggesting simultaneous 2$p$ emission.
Among the configurations occupied by the valence protons in $^{39}$Ti, the ${f_{7/2}}$ and ${p_{3/2}}$ orbitals warrant special attention. In our GCC analysis with the original interaction, the ${f_{7/2}}$ component dominates the configuration mixing with a $92\%$ share, contrasting with GSM configurations. This discrepancy likely arises because the core $^{37}$Ca is considered closed-shell for protons, complicating the consideration of core excitation and proton-neutron residual interactions within the GCC framework. 
To estimate the half-life uncertainty due to different configurations, we enhanced the  {potential depth} of the $p$ orbital to prioritize the $p$-wave component, as indicated by the black solid line in Fig.~\ref{fig_39Tidecaywidth}. Increasing the ${p_{3/2}}$ component significantly boosts the 2$p$ decay width and the likelihood of 2$p$ decay. Given the aforementioned configurations, we estimated the 2$p$ half-life limits at the calculated decay energy to range between 0.4 and 10 ms, closely aligning with $^{39}$Ti half-lives measured in previous experiments, as shown in the shaded band in Fig.~\ref{fig_39Tidecaywidth}.
Moreover, the comparable half-lives of 2$p$ and $\beta$ decay suggest a potential competitive relationship between these decay modes. When considering the impact of ${S_{2p}}$ uncertainty on half-life, the 2$p$ decay lifetime could extend beyond that of $\beta$ decay, potentially resulting in the non-observation of 2$p$ decay. Therefore, the as-yet-undiscovered 2$p$ emission in $^{39}$Ti could be attributed to its intense competition with other decay modes, primarily $\beta$ decay.

\begin{figure}[htb]
	\centering 
	\includegraphics[width=1\columnwidth, angle=0]{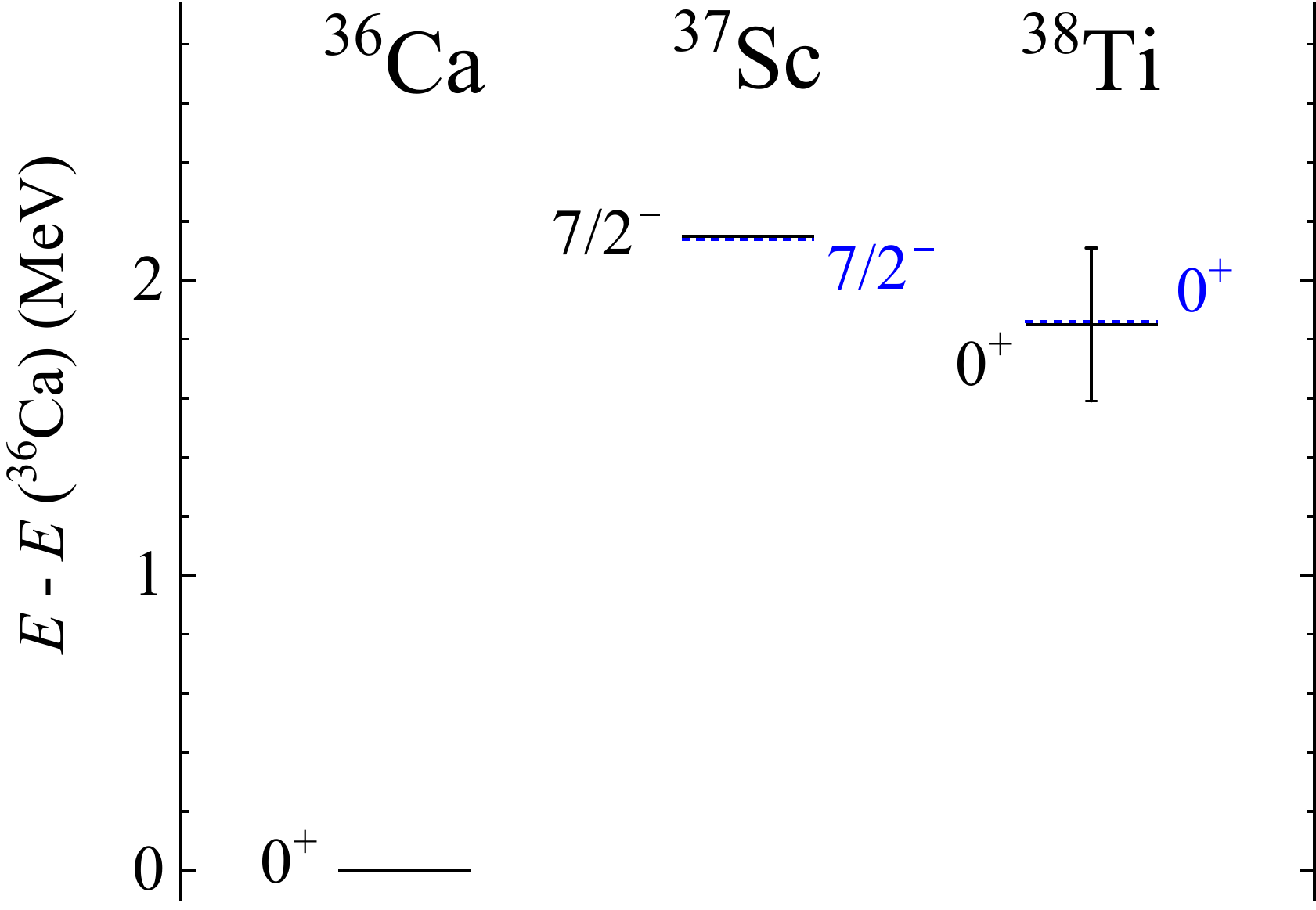}	
	\caption{Predicted spectra of $^{37}$Sc and $^{38}$Ti with respect to the inner core $^{36}$Ca using the optimized GSM Hamiltonian.  {As a comparision, the results with the channel-independent WS potential are calculated as shown by the blue dashed lines.}} 
	\label{fig_38Ti_energy}%
\end{figure}

To enhance our understanding of the internal structure of $^{39}$Ti, the 2$p$ density distributions within the g.s. were calculated for various configurations using Jacobi-$T$ coordinates through the GCC method. The resulting configurations are displayed in Fig.~\ref{fig_39Ti_2$p$density}. Panel (a) illustrates the results using the original interaction, while panels (b) and (c) depict scenarios in which the orbital strength of the single-particle WS potential for the $p$-wave and $f$-wave components has been adjusted. These adjustments aim to align the configuration with those obtained through the Gamow Shell Model (GSM) and to enhance the dominance of the $p$-wave, respectively.
As demonstrated in Fig.~\ref{fig_39Ti_2$p$density} (a), in the configuration where $f$-wave partial waves predominate, the two valence protons are most likely to pair, albeit with a minor probability of forming either a triangular or a cigar-like structure. Figure~\ref{fig_39Ti_2$p$density} (b) presents the 2$p$ density for $^{39}$Ti in a configuration akin to those previously obtained by the GSM. It is evident that as the occupation probability of the $p$-wave component increases, the likelihood of forming a triangular structure diminishes. When the $\ell = 1$ component becomes predominant, as seen in panel (c), the g.s. configuration of $^{39}$Ti is inclined to form both a diproton and a cigar-like structure.
Upon the analysis of the 2$p$ density distributions of $^{39}$Ti in these three configurations, it is evident that regardless of the specific configuration, the two valence protons exhibit a propensity to pair, which is mainly due to the overall attractive nuclear interaction.

As an isotope next to $^{39}$Ti, $^{38}$Ti, which is more proton-rich, has also attracted significant attention as a potential candidate for 2$p$ emission. In pursuit of understanding its 2$p$ radioactivity, both the GSM and the GCC methods were employed, analogous to the studies conducted on $^{39}$Ti. The energy spectra for $^{38}$Ti  {and} its neighboring isotones $^{37}$Sc and $^{36}$Ca, obtain from the GSM method, are illustrated in Fig.~\ref{fig_38Ti_energy}. 
The g.s. energy of $^{38}$Ti, denoted as two-proton decay energy ${Q_{2p}}$ with respect to $^{36}$Ca, is estimated to be approximately 1.85 MeV, with an uncertainty of 0.26 MeV. Meanwhile, the g.s. of $^{37}$Sc remains unidentified experimental; however, our results suggest a $7/2^-$ g.s., aligning with the mirror nucleus $^{37}$S~\cite{NNDC}, with a predicted g.s. energy of 2.15\,MeV. The single-proton separation energy, ${S_{p}}$ for $^{38}$Ti is calculated to be approximately 0.3 MeV,  {with an uncertainty of 0.26 MeV}. Despite the uncertainty in the g.s. energy of $^{38}$Ti, ${S_{p}}$ remains positive, indicating a higher likelihood of observing 2$p$ emission events in $^{38}$Ti compared to $^{39}$Ti. {Similar to $^{39}$Ti, the uncertainties of $^{37}$Sc and $^{38}$Ti caused by the single-particle potentials are also estimated. As shown in Fig.~\ref{fig_38Ti_energy}, the difference in the g.s. energies of $^{37}$Sc and $^{38}$Ti under the two sets of core-nucleon interactions is less than 0.01 MeV, which practically has no impact on the discussion of the results presented above.}

Given the similarity in the functional relationship between the 2$p$ half-life and decay energy under the GCC framework for both $^{39}$Ti and $^{38}$Ti, the 2$p$ half-life of $^{38}$Ti can be inferred from Fig.~\ref{fig_39Tidecaywidth}. Based on the predicted 2$p$ decay energy and the associated uncertainty, the 2$p$ half-life for $^{38}$Ti is estimated to range from $2.76 \times 10^{-14}$ seconds to $1.17 \times 10^{-11}$ seconds. Our predictions for the 2$p$ decay energy and half-life of $^{38}$Ti closely align with findings from other studies~\cite{ZHANG2023a,Mehana2023a,Santhosh2021a}.
Analysis of the g.s. configuration of $^{38}$Ti reveals that the two valence protons predominantly occupy the ${p_{3/2}}$ and ${f_{7/2}}$ orbitals. According to GSM predictions, the probability of the two valence protons in the g.s. of $^{38}$Ti occupying the $f$ orbital is 43\%, while the probability for the $p$ orbital is 41\%, indicating a tight competition between the $f$-wave and $p$-wave components in $^{38}$Ti. The g.s. of $^{38}$Ti contains a higher proportion of $p$-wave components compared to $^{39}$Ti, and given that the $p$-wave has a lower centrifugal barrier, it broadens the decay width and facilitates 2$p$ decay for $^{38}$Ti. Therefore, $^{38}$Ti is predicted to be a more promising candidate for 2$p$ emission than $^{39}$Ti.

\section{Summary}

The GSM employing an $A$-dependent EFT interaction has been utilized to study the spectra of $^{39}$Ti, a potential candidate for 2$p$ emission. The calculated 2$p$ separation energy for $^{39}$Ti suggests the possible observation of its 2$p$ decay phenomenon. Given the calculated energy uncertainties, the 2$p$ decay of $^{39}$Ti may occur as either a true 2$p$ decay or through a sequential emission process. Furthermore, the decay width of $^{39}$Ti has been analyzed as a function of the 2$p$ decay energy, contributing to an assessment of the half-life uncertainty, which is contingent upon the predominance of the $p$-wave and $f$-wave components. The 2$p$ half-life predicted by GCC method indicates a potential competition between the 2$p$ decay and $\beta$ decay processes in $^{39}$Ti. Additionally, due to the uncertainties associated with $S_{2p}$, the lifetime of the 2$p$ decay could potentially exceed that of $\beta$ decay, leading to a scenario where the 2$p$ decay remains undetected. Moreover, the 2$p$ density distributions of $^{39}$Ti across various configurations were systematically examined. It was observed that regardless of the configuration, the two valence protons exhibit a tendency to form proton  {pairs}.

The isotope $^{38}$Ti, akin to $^{39}$Ti, is also a candidate for 2$p$ decay and has been subjected to similar investigations. The 2$p$ decay energy for $^{38}$Ti is estimated to be approximately 1.85\,MeV, with an uncertainty of 0.26 MeV, while the single proton decay energy is around $-0.3$ MeV, with an uncertainty of 0.26\,MeV. These values indicate that the g.s. of $^{38}$Ti satisfies the energetic criteria for true 2$p$ decay. Employing a functional relationship between half-life and decay energy, analogous to that observed for $^{39}$Ti, the 2$p$ half-life for $^{38}$Ti is estimated to range from $2.76 \times 10^{-14}$ seconds to $1.17 \times 10^{-11}$ seconds. Relative to $^{39}$Ti, $^{38}$Ti exhibits a higher 2$p$ separation energy and a broader 2$p$ decay width, which facilitate the occurrence of 2$p$ emission.  {The present work motivates experimental searches for 2$p$ decay in $^{38,39}$Ti.} This prospective validation will enhance our understanding of 2$p$ decay processes and the nuclear structure characteristics of these isotopes.

\section*{Acknowledgements}

The authors thank Nicolas Michel and Marek P\l{}oszajczak for their meaningful discussion. This work is partially supported by the National Key Research and Development Program of China (Nos. 2023YFA1606404 and 2022YFA1602303), the National Natural Science Foundation of China (Nos. 12347106, 12147101, 12205340, 11925502, 11935001 and 11961141003), the Strategic Priority Research Program of Chinese Academy of Sciences (No. XDB34030000) and the Gansu Natural Science Foundation (No. 22JR5RA123).

\appendix

\bibliographystyle{elsarticle-num}

\begin{thebibliography}{10}
\expandafter\ifx\csname url\endcsname\relax
  \def\url#1{\texttt{#1}}\fi
\expandafter\ifx\csname urlprefix\endcsname\relax\def\urlprefix{URL }\fi
\expandafter\ifx\csname href\endcsname\relax
  \def\href#1#2{#2} \def\path#1{#1}\fi

\bibitem{GOLDANSKY1960a}
V.~I. Goldansky, On neutron-deficient isotopes of light nuclei and the phenomena of proton and two-proton radioactivity, Nucl. Phys. 19 (1960) 482--495.
\newblock \href {https://doi.org/10.1016/0029-5582(60)90258-3} {\path{doi:10.1016/0029-5582(60)90258-3}}.

\bibitem{GOLDANSKY1961a}
V.~I. Goldansky, Two-proton radioactivity, Nucl. Phys. 27 (1961) 648--664.
\newblock \href {https://doi.org/10.1007/BF02860176} {\path{doi:10.1007/BF02860176}}.

\bibitem{GALITSKY1964a}
V.~M. Galitsky, V.~F. Cheltsov, Two-proton radioactivity theory, Nucl. Phys. 56 (1964) 86--96.
\newblock \href {https://doi.org/10.1016/0029-5582(64)90455-9} {\path{doi:10.1016/0029-5582(64)90455-9}}.

\bibitem{JANECKE1965a}
J.~J${\rm{\ddot a}}$necke, The emission of protons from light neutron-deficient nuclei, Nucl. Phys. 61 (1965) 326--341.
\newblock \href {https://doi.org/10.1016/0029-5582(65)90907-7} {\path{doi:10.1016/0029-5582(65)90907-7}}.

\bibitem{Brown1991a}
B.~A. Brown, Diproton decay of nuclei on the proton drip line, Phys. Rev. C 43 (1991) R1513--R1517.
\newblock \href {https://doi.org/10.1103/PhysRevC.43.R1513} {\path{doi:10.1103/PhysRevC.43.R1513}}.

\bibitem{Brown1991b}
B.~A. Brown, {Erratum: Diproton decay of nuclei on the proton drip line}, Phys. Rev. C 44 (1991) 924--924.
\newblock \href {https://doi.org/10.1103/PhysRevC.44.924} {\path{doi:10.1103/PhysRevC.44.924}}.

\bibitem{Cole1996a}
B.~J. Cole, Stability of proton-rich nuclei in the upper $\mathit{sd}$ shell and lower $\mathit{pf}$ shell, Phys. Rev. C 54 (1996) 1240--1248.
\newblock \href {https://doi.org/10.1103/PhysRevC.54.1240} {\path{doi:10.1103/PhysRevC.54.1240}}.

\bibitem{Ormand1996a}
W.~E. Ormand, Properties of proton drip-line nuclei at the $sd$-$fp$-shell interface, Phys. Rev. C 53 (1996) 214--221.
\newblock \href {https://doi.org/10.1103/PhysRevC.53.214} {\path{doi:10.1103/PhysRevC.53.214}}.

\bibitem{Geesaman1977a}
D.~F. Geesaman, R.~L. McGrath, P.~M.~S. Lesser, P.~P. Urone, B.~VerWest, Particle decay of $^{6}\mathrm{Be}$, Phys. Rev. C 15 (1977) 1835--1838.
\newblock \href {https://doi.org/10.1103/PhysRevC.15.1835} {\path{doi:10.1103/PhysRevC.15.1835}}.

\bibitem{BOCHKAREV1989a}
O.~V. Bochkarev, L.~V. Chulkov, A.~A. Korsheninniicov, E.~A. Kuz'min, I.~G. Mukha, G.~B. Yankov, Democratic decay of $^{6}\mathrm{Be}$ states, Nucl. Phys. A 505 (1989) 215--240.
\newblock \href {https://doi.org/10.1016/0375-9474(89)90371-0} {\path{doi:10.1016/0375-9474(89)90371-0}}.

\bibitem{KeKelis1978a}
G.~J. KeKelis, M.~S. Zisman, D.~K. Scott, R.~Jahn, D.~J. Vieira, J.~Cerny, F.~Ajzenberg-Selove, Masses of the unbound nuclei $^{16}\mathrm{Ne}$, $^{15}\mathrm{F}$, and $^{12}\mathrm{O}$, Phys. Rev. C 17 (1978) 1929--1938.
\newblock \href {https://doi.org/10.1103/PhysRevC.17.1929} {\path{doi:10.1103/PhysRevC.17.1929}}.

\bibitem{Kryger1995a}
R.~A. Kryger, A.~Azhari, M.~Hellstr\"om, J.~H. Kelley, T.~Kubo, R.~Pfaff, E.~Ramakrishnan, B.~M. Sherrill, M.~Thoennessen, S.~Yokoyama, R.~J. Charity, J.~Dempsey, A.~Kirov, N.~Robertson, D.~G. Sarantites, L.~G. Sobotka, J.~A. Winger, {Two-Proton Emission from the Ground State of $^{12}\mathrm{O}$}, Phys. Rev. Lett. 74 (1995) 860--863.
\newblock \href {https://doi.org/10.1103/PhysRevLett.74.860} {\path{doi:10.1103/PhysRevLett.74.860}}.

\bibitem{Mukha2007a}
I.~Mukha, K.~S\"ummerer, L.~Acosta, M.~A.~G. Alvarez, E.~Casarejos, A.~Chatillon, D.~Cortina-Gil, J.~Espino, A.~Fomichev, J.~E. Garc\'{\i}a-Ramos, H.~Geissel, J.~G\'omez-Camacho, L.~Grigorenko, J.~Hoffmann, O.~Kiselev, A.~Korsheninnikov, N.~Kurz, Y.~Litvinov, I.~Martel, C.~Nociforo, W.~Ott, M.~Pf\"utzner, C.~Rodr\'{\i}guez-Tajes, E.~Roeckl, M.~Stanoiu, H.~Weick, P.~J. Woods, {Observation of Two-Proton Radioactivity of $^{19}\mathrm{Mg}$ by Tracking the Decay Products}, Phys. Rev. Lett. 99 (2007) 182501.
\newblock \href {https://doi.org/10.1103/PhysRevLett.99.182501} {\path{doi:10.1103/PhysRevLett.99.182501}}.

\bibitem{Mukha2008a}
I.~Mukha, L.~Grigorenko, K.~S\"ummerer, L.~Acosta, M.~A.~G. Alvarez, E.~Casarejos, A.~Chatillon, D.~Cortina-Gil, J.~M. Espino, A.~Fomichev, J.~E. Garc\'{\i}a-Ramos, H.~Geissel, J.~G\'omez-Camacho, J.~Hofmann, O.~Kiselev, A.~Korsheninnikov, N.~Kurz, Y.~Litvinov, I.~Martel, C.~Nociforo, W.~Ott, M.~Pf\"utzner, C.~Rodr\'{\i}guez-Tajes, E.~Roeckl, M.~Stanoiu, H.~Weick, P.~J. Woods, Proton-proton correlations observed in two-proton decay of $^{19}\mathrm{Mg}$ and $^{16}\mathrm{Ne}$, Phys. Rev. C 77 (2008) 061303.
\newblock \href {https://doi.org/10.1103/PhysRevC.77.061303} {\path{doi:10.1103/PhysRevC.77.061303}}.

\bibitem{Giovinazzo2002a}
J.~Giovinazzo, B.~Blank, M.~Chartier, S.~Czajkowski, A.~Fleury, M.~J. Lopez~Jimenez, M.~S. Pravikoff, J.-C. Thomas, F.~de~Oliveira~Santos, M.~Lewitowicz, V.~Maslov, M.~Stanoiu, R.~Grzywacz, M.~Pf\"utzner, C.~Borcea, B.~A. Brown, {Two-Proton Radioactivity of $^{\mathrm{45}}\mathrm{F}\mathrm{e}$}, Phys. Rev. Lett. 89 (2002) 102501.
\newblock \href {https://doi.org/10.1103/PhysRevLett.89.102501} {\path{doi:10.1103/PhysRevLett.89.102501}}.

\bibitem{Pfutzner2002a}
M.~Pf\"utzner, E.~Badura, C.~Bingham, B.~Blank, M.~Chartier, H.~Geissel, J.~Giovinazzo, L.~V. Grigorenko, R.~Grzywacz, M.~Hellström, Z.~Janas, J.~Kurcewicz, A.~S. Lalleman, C.~Mazzocchi, I.~Mukha, G.~M\"unzenberg, C.~Plettner, E.~Roeckl, K.~P. Rykaczewski, K.~Schmidt, R.~S. Simon, M.~Stanoiu, J.~C. Thomas, First evidence for the two-proton decay of $^{45}\mathrm{Fe}$, Eur. Phys. J. A 14 (2002) 279--285.
\newblock \href {https://doi.org/10.1140/epja/i2002-10033-9} {\path{doi:10.1140/epja/i2002-10033-9}}.

\bibitem{Blank2005a}
B.~Blank, A.~Bey, G.~Canchel, C.~Dossat, A.~Fleury, J.~Giovinazzo, I.~Matea, N.~Adimi, F.~De~Oliveira, I.~Stefan, G.~Georgiev, S.~Gr\'evy, J.~C. Thomas, C.~Borcea, D.~Cortina, M.~Caamano, M.~Stanoiu, F.~Aksouh, B.~A. Brown, F.~C. Barker, W.~A. Richter, {First Observation of $^{54}\mathrm{Zn}$ and its Decay by Two-Proton Emission}, Phys. Rev. Lett. 94 (2005) 232501.
\newblock \href {https://doi.org/10.1103/PhysRevLett.94.232501} {\path{doi:10.1103/PhysRevLett.94.232501}}.

\bibitem{Ascher2011a}
P.~Ascher, L.~Audirac, N.~Adimi, B.~Blank, C.~Borcea, B.~A. Brown, I.~Companis, F.~Delalee, C.~E. Demonchy, F.~de~Oliveira~Santos, J.~Giovinazzo, S.~Gr\'evy, L.~V. Grigorenko, T.~Kurtukian-Nieto, S.~Leblanc, J.-L. Pedroza, L.~Perrot, J.~Pibernat, L.~Serani, P.~C. Srivastava, J.-C. Thomas, {Direct Observation of Two Protons in the Decay of $^{54}\mathrm{Zn}$}, Phys. Rev. Lett. 107 (2011) 102502.
\newblock \href {https://doi.org/10.1103/PhysRevLett.107.102502} {\path{doi:10.1103/PhysRevLett.107.102502}}.

\bibitem{Dossat2005a}
C.~Dossat, A.~Bey, B.~Blank, G.~Canchel, A.~Fleury, J.~Giovinazzo, I.~Matea, F.~d.~O. Santos, G.~Georgiev, S.~Gr\'evy, I.~Stefan, J.~C. Thomas, N.~Adimi, C.~Borcea, D.~C. Gil, M.~Caamano, M.~Stanoiu, F.~Aksouh, B.~A. Brown, L.~V. Grigorenko, Two-proton radioactivity studies with $^{45}\mathrm{Fe}$ and $^{48}\mathrm{Ni}$, Phys. Rev. C 72 (2005) 054315.
\newblock \href {https://doi.org/10.1103/PhysRevC.72.054315} {\path{doi:10.1103/PhysRevC.72.054315}}.

\bibitem{Pomorski2011a}
M.~Pomorski, M.~Pf\"utzner, W.~Dominik, R.~Grzywacz, T.~Baumann, J.~S. Berryman, H.~Czyrkowski, R.~D\k{a}browski, T.~Ginter, J.~Johnson, G.~Kami\ifmmode~\acute{n}\else \'{n}\fi{}ski, A.~Ku\ifmmode~\acute{z}\else \'{z}\fi{}niak, N.~Larson, S.~N. Liddick, M.~Madurga, C.~Mazzocchi, S.~Mianowski, K.~Miernik, D.~Miller, S.~Paulauskas, J.~Pereira, K.~P. Rykaczewski, A.~Stolz, S.~Suchyta, First observation of two-proton radioactivity in $^{48}\mathrm{Ni}$, Phys. Rev. C 83 (2011) 061303.
\newblock \href {https://doi.org/10.1103/PhysRevC.83.061303} {\path{doi:10.1103/PhysRevC.83.061303}}.

\bibitem{Goigoux2016a}
T.~Goigoux, P.~Ascher, B.~Blank, M.~Gerbaux, J.~Giovinazzo, S.~Gr\'evy, T.~Kurtukian~Nieto, C.~Magron, P.~Doornenbal, G.~G. Kiss, S.~Nishimura, P.~A. S\"oderstr\"om, V.~H. Phong, J.~Wu, D.~S. Ahn, N.~Fukuda, N.~Inabe, T.~Kubo, S.~Kubono, H.~Sakurai, Y.~Shimizu, T.~Sumikama, H.~Suzuki, H.~Takeda, J.~Agramunt, A.~Algora, V.~Guadilla, A.~Montaner-Piza, A.~I. Morales, S.~E.~A. Orrigo, B.~Rubio, Y.~Fujita, M.~Tanaka, W.~Gelletly, P.~Aguilera, F.~Molina, F.~Diel, D.~Lubos, G.~de~Angelis, D.~Napoli, C.~Borcea, A.~Boso, R.~B. Cakirli, E.~Ganioglu, J.~Chiba, D.~Nishimura, H.~Oikawa, Y.~Takei, S.~Yagi, K.~Wimmer, G.~de~France, S.~Go, B.~A. Brown, {Two-Proton Radioactivity of $^{67}\mathrm{Kr}$}, Phys. Rev. Lett. 117 (2016) 162501.
\newblock \href {https://doi.org/10.1103/PhysRevLett.117.162501} {\path{doi:10.1103/PhysRevLett.117.162501}}.

\bibitem{Grigorenko2017a}
L.~V. Grigorenko, T.~A. Golubkova, J.~S. Vaagen, M.~V. Zhukov, Decay mechanism and lifetime of $^{67}\mathrm{Kr}$, Phys. Rev. C 95 (2017) 021601.
\newblock \href {https://doi.org/10.1103/PhysRevC.95.021601} {\path{doi:10.1103/PhysRevC.95.021601}}.

\bibitem{Wang2018a}
S.~M. Wang, W.~Nazarewicz, {Puzzling Two-Proton Decay of $^{67}\mathrm{Kr}$}, Phys. Rev. Lett. 120 (2018) 212502.
\newblock \href {https://doi.org/10.1103/PhysRevLett.120.212502} {\path{doi:10.1103/PhysRevLett.120.212502}}.

\bibitem{Pfutzner2012a}
M.~Pf\"utzner, M.~Karny, L.~V. Grigorenko, K.~Riisager, Radioactive decays at limits of nuclear stability, Rev. Mod. Phys. 84 (2012) 567--619.
\newblock \href {https://doi.org/10.1103/RevModPhys.84.567} {\path{doi:10.1103/RevModPhys.84.567}}.

\bibitem{Blank2008a}
B.~Blank, M.~P\l{}oszajczak, Two-proton radioactivity, Rep. Prog. Phys. 71 (2008) 046301.
\newblock \href {https://doi.org/10.1088/0034-4885/71/4/046301} {\path{doi:10.1088/0034-4885/71/4/046301}}.

\bibitem{Grigorenko2009a}
L.~V. Grigorenko, Theoretical study of two-proton radioactivity. status, predictions, and applications, Phys. Part. Nucl. 40 (2009).
\newblock \href {https://doi.org/10.1134/S1063779609050049} {\path{doi:10.1134/S1063779609050049}}.

\bibitem{Zhou2022a}
L.~Zhou, S.~M. Wang, D.~Q. Fang, Y.~G. Ma, {Recent Progress in Two-proton Radioactivity}, Nucl. Sci. Tech. 33 (2022).
\newblock \href {https://doi.org/10.1007/s41365-022-01091-1} {\path{doi:10.1007/s41365-022-01091-1}}.

\bibitem{Pfutzner2023a}
M.~Pf\"utzner, I.~Mukha, S.~M. Wang, Two-proton emission and related phenomena, Prog. Part. Nucl. Phys. 132 (2023) 104050.
\newblock \href {https://doi.org/10.1016/j.ppnp.2023.104050} {\path{doi:10.1016/j.ppnp.2023.104050}}.

\bibitem{GOLDANSKII1988a}
V.~I. Goldanskii, Neutron-excessive nuclei and two-proton radioactivity, Phys. Lett. B 212 (1988) 11--12.
\newblock \href {https://doi.org/10.1016/0370-2693(88)91226-9} {\path{doi:10.1016/0370-2693(88)91226-9}}.

\bibitem{DETRAZ1990a}
C.~D\'{e}traz, R.~Anne, P.~Bricault, D.~Guillemaud-Mueller, M.~Lewitowicz, A.~C. Mueller, Z.~Yu~Hu, V.~Borrel, J.~C. Jacmart, F.~Pougheon, A.~Richard, D.~Bazin, J.~P. Dufour, A.~Fleury, F.~Hubert, M.~S. Pravikoff, {Search for direct two-proton radioactivity from Ti isotopes at the proton drip line}, Nucl. Phys. A 519 (1990) 529--547.
\newblock \href {https://doi.org/10.1016/0375-9474(90)90445-R} {\path{doi:10.1016/0375-9474(90)90445-R}}.

\bibitem{Moltz1992a}
D.~Moltz, J.~C. Batchelder, T.~F. Lang, T.~J. Ognibene, J.~Cerny, P.~E. Haustein, P.~L. de~Reeder, Beta-delayed two-proton decay of $^{39}\mathrm{Ti}$, Z. Phys. A 342 (1992) 273--276.
\newblock \href {https://doi.org/10.1007/BF01291509} {\path{doi:10.1007/BF01291509}}.

\bibitem{Giovinazzo2001a}
J.~Giovinazzo, B.~Blank, C.~Borcea, M.~Chartier, S.~Czajkowski, G.~de~France, R.~Grzywacz, Z.~Janas, M.~Lewitowicz, F.~de~Oliveira~Santos, M.~Pf\"utzner, M.~S. Pravikoff, J.~C. Thomas, Decay of proton-rich nuclei between $^{39}\mathrm{Ti}$ and $^{49}\mathrm{Ni}$, Eur. Phys. J. A 10 (2001) 73--84.
\newblock \href {https://doi.org/10.1007/s100500170146} {\path{doi:10.1007/s100500170146}}.

\bibitem{DOSSAT2007a}
C.~Dossat, N.~Adimi, F.~Aksouh, F.~Becker, A.~Bey, B.~Blank, C.~Borcea, R.~Borcea, A.~Boston, M.~Caamano, G.~Canchel, M.~Chartier, D.~Cortina, S.~Czajkowski, G.~de~France, F.~de~Oliveira~Santos, A.~Fleury, G.~Georgiev, J.~Giovinazzo, S.~Gr\'{e}vy, R.~Grzywacz, M.~Hellström, M.~Honma, Z.~Janas, D.~Karamanis, J.~Kurcewicz, M.~Lewitowicz, M.~J. L\'{o}pez~Jim\'{e}nez, C.~Mazzocchi, I.~Matea, V.~Maslov, P.~Mayet, C.~Moore, M.~Pf\"utzner, M.~S. Pravikoff, M.~Stanoiu, I.~Stefan, J.~C. Thomas, {The decay of proton-rich nuclei in the mass A=36–56 region}, Nucl. Phys. A 792 (2007) 18--86.
\newblock \href {https://doi.org/10.1016/j.nuclphysa.2007.05.004} {\path{doi:10.1016/j.nuclphysa.2007.05.004}}.

\bibitem{Cole1997a}
B.~J. Cole, Systematics of proton and diproton separation energies for light nuclei, Phys. Rev. C 56 (1997) 1866--1871.
\newblock \href {https://doi.org/10.1103/PhysRevC.56.1866} {\path{doi:10.1103/PhysRevC.56.1866}}.

\bibitem{Grigorenko2001a}
L.~V. Grigorenko, R.~C. Johnson, I.~G. Mukha, I.~J. Thompson, M.~V. Zhukov, Two-proton radioactivity and three-body decay: General problems and theoretical approach, Phys. Rev. C 64 (2001) 054002.
\newblock \href {https://doi.org/10.1103/PhysRevC.64.054002} {\path{doi:10.1103/PhysRevC.64.054002}}.

\bibitem{Tian2013a}
J.~Tian, N.~Wang, C.~Li, J.~Li, {Improved Kelson-Garvey mass relations for proton-rich nuclei}, Phys. Rev. C 87 (2013) 014313.
\newblock \href {https://doi.org/10.1103/PhysRevC.87.014313} {\path{doi:10.1103/PhysRevC.87.014313}}.

\bibitem{MC2020a}
C.~Ma, Y.~Y. Zong, Y.~M. Zhao, A.~Arima, Mass relations of mirror nuclei with local correlations, Phys. Rev. C 102 (2020) 024330.
\newblock \href {https://doi.org/10.1103/PhysRevC.102.024330} {\path{doi:10.1103/PhysRevC.102.024330}}.

\bibitem{ZHANG2023a}
Z.~Zhang, C.~Yuan, C.~Qi, B.~Cai, X.~Xu, {Extended R-matrix description of two-proton radioactivity}, Phys. Lett. B 838 (2023) 137740.
\newblock \href {https://doi.org/10.1016/j.physletb.2023.137740} {\path{doi:10.1016/j.physletb.2023.137740}}.

\bibitem{Cui2020a}
J.~P. Cui, Y.~H. Gao, Y.~Z. Wang, J.~Z. Gu, Two-proton radioactivity within a generalized liquid drop model, Phys. Rev. C 101 (2020) 014301.
\newblock \href {https://doi.org/10.1103/PhysRevC.101.014301} {\path{doi:10.1103/PhysRevC.101.014301}}.

\bibitem{Royer2022a}
G.~Royer, Calculation of two-proton radioactivity and application to $^{9}\mathrm{Be},$ $^{6,7}\mathrm{Li},$ $^{3,6}\mathrm{He},$ and $^{2,3}\mathrm{H}$ emissions, Phys. Rev. C 106 (2022) 034605.
\newblock \href {https://doi.org/10.1103/PhysRevC.106.034605} {\path{doi:10.1103/PhysRevC.106.034605}}.

\bibitem{Santhosh2021a}
K.~P. Santhosh, Theoretical studies on two-proton radioactivity, Phys. Rev. C 104 (2021) 064613.
\newblock \href {https://doi.org/10.1103/PhysRevC.104.064613} {\path{doi:10.1103/PhysRevC.104.064613}}.

\bibitem{Santhosh2022a}
K.~P. Santhosh, {Two-proton radioactivity within a Coulomb and proximity potential model for deformed nuclei}, Phys. Rev. C 106 (2022) 054604.
\newblock \href {https://doi.org/10.1103/PhysRevC.106.054604} {\path{doi:10.1103/PhysRevC.106.054604}}.

\bibitem{Neufcourt2020a}
L.~Neufcourt, Y.~Cao, S.~Giuliani, W.~Nazarewicz, E.~Olsen, O.~B. Tarasov, {Beyond the proton drip line: Bayesian analysis of proton-emitting nuclei}, Phys. Rev. C 101 (2020) 014319.
\newblock \href {https://doi.org/10.1103/PhysRevC.101.014319} {\path{doi:10.1103/PhysRevC.101.014319}}.

\bibitem{Mehana2023a}
P.~Mehana, N.~S. Rajeswari, Two-proton and one-proton emission of two-proton emitters, Eur. Phys. J. A 59 (2023) 104.
\newblock \href {https://doi.org/10.1140/epja/s10050-023-01004-9} {\path{doi:10.1140/epja/s10050-023-01004-9}}.

\bibitem{XU2006a}
F.~R. Xu, J.~C. Pei, Mean-field cluster potentials for various cluster decays, Phys. Lett. B 642 (2006) 322--325.
\newblock \href {https://doi.org/10.1016/j.physletb.2006.09.048} {\path{doi:10.1016/j.physletb.2006.09.048}}.

\bibitem{Okolowicz2012a}
J.~Oko\l{}owicz, M.~Pf\"utzner, W.~Nazarewicz, {On the Origin of Nuclear Clustering}, Prog. theor. phys., Suppl 196 (2012) 230--243.
\newblock \href {https://doi.org/10.1143/PTPS.196.230} {\path{doi:10.1143/PTPS.196.230}}.

\bibitem{Michel_2021b}
N.~Michel, M.~P\l{}oszajczak, The Gamow Shell Model: the unified theory of nuclear structure and reactions, Springer Berlin Heidelberg, Berlin, Heidelberg, 2021.
\newblock \href {https://doi.org/10.1007/978-3-030-69356-5} {\path{doi:10.1007/978-3-030-69356-5}}.

\bibitem{Bennaceur2000a}
K.~Bennaceur, F.~Nowacki, J.~Oko\l{}owicz, M.~P\l{}oszajczak, Analysis of the $^{16}\mathrm{O}(p,\gamma )^{17}\mathrm{F}$ capture reaction using the shell model embedded in the continuum, Nucl. Phys. A 671~(1) (2000) 203--232.
\newblock \href {https://doi.org/10.1016/S0375-9474(99)00851-9} {\path{doi:10.1016/S0375-9474(99)00851-9}}.

\bibitem{Okolowicz2003a}
J.~Oko\l{}owicz, M.~P\l{}oszajczak, I.~Rotter, Dynamics of quantum systems embedded in a continuum, Phys. Rep 374~(4) (2003) 271--383.
\newblock \href {https://doi.org/10.1016/S0370-1573(02)00366-6} {\path{doi:10.1016/S0370-1573(02)00366-6}}.

\bibitem{Betan2002a}
R.~Id~Betan, R.~J. Liotta, N.~Sandulescu, T.~Vertse, {Two-Particle Resonant States in a Many-Body Mean Field}, Phys. Rev. Lett. 89 (2002) 042501.
\newblock \href {https://doi.org/10.1103/PhysRevLett.89.042501} {\path{doi:10.1103/PhysRevLett.89.042501}}.

\bibitem{Michel2002a}
N.~Michel, W.~Nazarewicz, M.~P\l{}oszajczak, K.~Bennaceur, {Gamow Shell Model Description of Neutron-Rich Nuclei}, Phys. Rev. Lett. 89 (2002) 042502.
\newblock \href {https://doi.org/10.1103/PhysRevLett.89.042502} {\path{doi:10.1103/PhysRevLett.89.042502}}.

\bibitem{Forssen2013a}
C.~Forss\'{e}n, G.~Hagen, M.~Hjorth-Jensen, W.~Nazarewicz, J.~Rotureau, Living on the edge of stability, the limits of the nuclear landscape, Phys. Scr. 2013 (2013) 014022.
\newblock \href {https://doi.org/10.1088/0031-8949/2013/T152/014022} {\path{doi:10.1088/0031-8949/2013/T152/014022}}.

\bibitem{Jaganathen2017a}
Y.~Jaganathen, R.~M.~I. Betan, N.~Michel, W.~Nazarewicz, M.~P\l{}oszajczak, {Quantified Gamow shell model interaction for $psd$-shell nuclei}, Phys. Rev. C 96 (2017) 054316.
\newblock \href {https://doi.org/10.1103/PhysRevC.96.054316} {\path{doi:10.1103/PhysRevC.96.054316}}.

\bibitem{Wang2017a}
S.~M. Wang, N.~Michel, W.~Nazarewicz, F.~R. Xu, {Structure and decays of nuclear three-body systems: The Gamow coupled-channel method in Jacobi coordinates}, Phys. Rev. C 96 (2017) 044307.
\newblock \href {https://doi.org/10.1103/PhysRevC.96.044307} {\path{doi:10.1103/PhysRevC.96.044307}}.

\bibitem{Rotureau2005a}
J.~Rotureau, J.~Oko\l{}owicz, M.~P\l{}oszajczak, Microscopic theory of the two-proton radioactivity, Phys. Rev. Lett. 95 (2005) 042503.
\newblock \href {https://doi.org/10.1103/PhysRevLett.95.042503} {\path{doi:10.1103/PhysRevLett.95.042503}}.

\bibitem{Michel2021a}
N.~Michel, J.~G. Li, F.~R. Xu, W.~Zuo, Proton decays in $^{16}\mathrm{Ne}$ and $^{18}\mathrm{Mg}$ and isospin-symmetry breaking in carbon isotopes and isotones, Phys. Rev. C 103 (2021) 044319.
\newblock \href {https://doi.org/10.1103/PhysRevC.103.044319} {\path{doi:10.1103/PhysRevC.103.044319}}.

\bibitem{Wang2019a}
S.~M. Wang, W.~Nazarewicz, R.~J. Charity, L.~G. Sobotka, Structure and decay of the extremely proton-rich nuclei $^{11,12}\mathrm{O}$, Phys. Rev. C 99 (2019) 054302.
\newblock \href {https://doi.org/10.1103/PhysRevC.99.054302} {\path{doi:10.1103/PhysRevC.99.054302}}.

\bibitem{Yang2023a}
Y.~H. Yang, Y.~G. Ma, S.~M. Wang, B.~Zhou, D.~Q. Fang, Structure and decay mechanism of the low-lying states in $^{9}\mathrm{Be}$ and $^{9}\mathrm{B}$, Phys. Rev. C 108 (2023) 044307.
\newblock \href {https://doi.org/10.1103/PhysRevC.108.044307} {\path{doi:10.1103/PhysRevC.108.044307}}.

\bibitem{Rotureau2006a}
J.~Rotureau, J.~Oko\l{}owicz, M.~P\l{}oszajczak, Theory of the two-proton radioactivity in the continuum shell model, Nucl. Phys. A 767 (2006) 13--57.
\newblock \href {https://doi.org/10.1016/j.nuclphysa.2005.12.005} {\path{doi:10.1016/j.nuclphysa.2005.12.005}}.

\bibitem{Wang2021a}
S.~M. Wang, W.~Nazarewicz, Fermion pair dynamics in open quantum systems, Phys. Rev. Lett. 126 (2021) 142501.
\newblock \href {https://doi.org/10.1103/PhysRevLett.126.142501} {\path{doi:10.1103/PhysRevLett.126.142501}}.

\bibitem{Wang2022a}
S.~M. Wang, W.~Nazarewicz, R.~J. Charity, L.~G. Sobotka, Nucleon–nucleon correlations in the extreme oxygen isotopes, J. Phys. G, Nucl. Part. Phys. 49 (2022) 10LT02.
\newblock \href {https://doi.org/10.1088/1361-6471/ac888f} {\path{doi:10.1088/1361-6471/ac888f}}.

\bibitem{SUN2017a}
Z.~H. Sun, Q.~Wu, Z.~H. Zhao, B.~S. Hu, S.~J. Dai, F.~R. Xu, {Resonance and continuum Gamow shell model with realistic nuclear forces}, Phys. Lett. B 769 (2017) 227--232.
\newblock \href {https://doi.org/10.1016/j.physletb.2017.03.054} {\path{doi:10.1016/j.physletb.2017.03.054}}.

\bibitem{BERGGREN1968a}
T.~Berggren, On the use of resonant states in eigenfunction expansions of scattering and reaction amplitudes, Nucl. Phys. A 109 (1968) 265--287.
\newblock \href {https://doi.org/10.1016/0375-9474(68)90593-9} {\path{doi:10.1016/0375-9474(68)90593-9}}.

\bibitem{MACHLEIDT2011a}
R.~Machleidt, D.~R. Entem, Chiral effective field theory and nuclear forces, Phys. Rep 503 (2011) 1--75.
\newblock \href {https://doi.org/10.1016/j.physrep.2011.02.001} {\path{doi:10.1016/j.physrep.2011.02.001}}.

\bibitem{CONTESSI2017a}
L.~Contessi, A.~Lovato, F.~Pederiva, A.~Roggero, J.~Kirscher, U.~van Kolck, {Ground-state properties of $^{4}\mathrm{He}$ and $^{16}\mathrm{O}$ extrapolated from lattice QCD with pionless EFT}, Phys. Lett. B 772 (2017) 839--848.
\newblock \href {https://doi.org/10.1016/j.physletb.2017.07.048} {\path{doi:10.1016/j.physletb.2017.07.048}}.

\bibitem{Hammer2013a}
H.-W. Hammer, A.~Nogga, A.~Schwenk, Colloquium: Three-body forces: From cold atoms to nuclei, Rev. Mod. Phys. 85 (2013) 197--217.
\newblock \href {https://doi.org/10.1103/RevModPhys.85.197} {\path{doi:10.1103/RevModPhys.85.197}}.

\bibitem{Hammer2020a}
H.-W. Hammer, S.~K\"onig, U.~van Kolck, Nuclear effective field theory: Status and perspectives, Rev. Mod. Phys. 92 (2020) 025004.
\newblock \href {https://doi.org/10.1103/RevModPhys.92.025004} {\path{doi:10.1103/RevModPhys.92.025004}}.

\bibitem{Michel2019a}
N.~Michel, J.~G. Li, F.~R. Xu, W.~Zuo, {Description of proton-rich nuclei in the $A\ensuremath{\approx}20$ region within the Gamow shell model}, Phys. Rev. C 100 (2019) 064303.
\newblock \href {https://doi.org/10.1103/PhysRevC.100.064303} {\path{doi:10.1103/PhysRevC.100.064303}}.

\bibitem{Brown2006a}
B.~A. Brown, W.~A. Richter, {New ``USD'' Hamiltonians for the $\mathit{sd}$ shell}, Phys. Rev. C 74 (2006) 034315.
\newblock \href {https://doi.org/10.1103/PhysRevC.74.034315} {\path{doi:10.1103/PhysRevC.74.034315}}.

\bibitem{Huth2018a}
L.~Huth, V.~Durant, J.~Simonis, A.~Schwenk, Shell-model interactions from chiral effective field theory, Phys. Rev. C 98 (2018) 044301.
\newblock \href {https://doi.org/10.1103/PhysRevC.98.044301} {\path{doi:10.1103/PhysRevC.98.044301}}.

\bibitem{Li2021b}
J.~G. Li, N.~Michel, W.~Zuo, F.~R. Xu, {Unbound spectra of neutron-rich oxygen isotopes predicted by the Gamow shell model}, Phys. Rev. C 103 (2021) 034305.
\newblock \href {https://doi.org/10.1103/PhysRevC.103.034305} {\path{doi:10.1103/PhysRevC.103.034305}}.

\bibitem{Binder2016a}
S.~Binder, A.~Ekstr\"om, G.~Hagen, T.~Papenbrock, K.~A. Wendt, Effective field theory in the harmonic oscillator basis, Phys. Rev. C 93 (2016) 044332.
\newblock \href {https://doi.org/10.1103/PhysRevC.98.054301} {\path{doi:10.1103/PhysRevC.98.054301}}.

\bibitem{Bansal2018a}
A.~Bansal, S.~Binder, A.~Ekstr\"om, G.~Hagen, G.~R. Jansen, T.~Papenbrock, Pion-less effective field theory for atomic nuclei and lattice nuclei, Phys. Rev. C 98 (2018) 054301.
\newblock \href {https://doi.org/10.1103/PhysRevC.98.054301} {\path{doi:10.1103/PhysRevC.98.054301}}.

\bibitem{SM}
See Supplemental Material at [URL inserted by publisher] for more details on the Effective Field Theory interaction, optimization, and Jacobi coordinates.

\bibitem{Dobaczewski2014a}
J.~Dobaczewski, W.~Nazarewicz, P.-G. Reinhard, Error estimates of theoretical models: a guide, J. Phys. G. Nucl. Part. Phys. 41~(7) (2014) 074001.
\newblock \href {https://doi.org/10.1088/0954-3899/41/7/074001} {\path{doi:10.1088/0954-3899/41/7/074001}}.

\bibitem{Cwiok1987a}
S.~Cwiok, J.~Dudek, W.~Nazarewicz, J.~Skalski, T.~Werner, Single-particle energies, wave functions, quadrupole moments and g-factors in an axially deformed woods-saxon potential with applications to the two-centre-type nuclear problems, Comput. Phys. Commun. 46~(3) (1987) 379--399.
\newblock \href {https://doi.org/10.1016/0010-4655(87)90093-2} {\path{doi:10.1016/0010-4655(87)90093-2}}.

\bibitem{Dudek1982}
J.~Dudek, Z.~Szyma\ifmmode~\acute{n}\else \'{n}\fi{}ski, T.~Werner, A.~Faessler, C.~Lima, \href{https://link.aps.org/doi/10.1103/PhysRevC.26.1712}{Description of the high spin states in $^{146}\mathrm{Gd}$ using the optimized woods-saxon potential}, Phys. Rev. C 26 (1982) 1712--1718.
\newblock \href {https://doi.org/10.1103/PhysRevC.26.1712} {\path{doi:10.1103/PhysRevC.26.1712}}.
\newline\urlprefix\url{https://link.aps.org/doi/10.1103/PhysRevC.26.1712}

\bibitem{Nazarewicz1985}
W.~Nazarewicz, J.~Dudek, R.~Bengtsson, T.~Bengtsson, T.~Ragnarsson, {Microscopic study of the high-spin behaviour in selected A $\approx$ 80 nuclei}, Nucl. Rhys. A 435 (1985) 397--447.
\newblock \href {https://doi.org/10.1016/0375-9474(85)90471-3} {\path{doi:10.1016/0375-9474(85)90471-3}}.

\bibitem{Thompson1977a}
D.~R. Thompson, M.~Lemere, Y.~C. Tang, Systematic investigation of scattering problems with the resonating-group method, Nucl. Phys. A 286 (1977) 53--66.
\newblock \href {https://doi.org/10.1016/0375-9474(77)90007-0} {\path{doi:10.1016/0375-9474(77)90007-0}}.

\bibitem{ENDT1978a}
P.~Endt, C.~{Van Der Leun}, Energy levels of a = 21–44 nuclei (vi), Nucl. Phys. A 310~(1) (1978) 1--751.
\newblock \href {https://doi.org/https://doi.org/10.1016/0375-9474(78)90611-5} {\path{doi:https://doi.org/10.1016/0375-9474(78)90611-5}}.

\bibitem{NNDC}
National nuclear data center, \url{http://www.nndc.bnl.gov/}, 2024 (accessed 31 August 2024).

\bibitem{Dronchi2023a}
N.~Dronchi, D.~Weisshaar, B.~A. Brown, A.~Gade, R.~J. Charity, L.~G. Sobotka, K.~W. Brown, W.~Reviol, D.~Bazin, P.~J. Farris, A.~M. Hill, J.~Li, B.~Longfellow, D.~Rhodes, S.~N. Paneru, S.~A. Gillespie, A.~Anthony, E.~Rubino, S.~Biswas, {Measurement of the $B(E2\ensuremath{\uparrow})$ strengths of $^{36}\mathrm{Ca}$ and $^{38}\mathrm{Ca}$}, Phys. Rev. C 107 (2023) 034306.
\newblock \href {https://doi.org/10.1103/PhysRevC.107.034306} {\path{doi:10.1103/PhysRevC.107.034306}}.

\end{thebibliography}






\end{document}